\documentclass[12pt]{article}
\usepackage{amssymb}

\newcommand{\bi}{\bibitem}
\newcommand{\be}{\begin{eqnarray}}
\newcommand{\ee}{\end{eqnarray}}
\newcommand{\rar}{\rightarrow}

\begin{document}


\begin{titlepage}

\title{
On the Possibility of Super-luminal Propagation in a Gravitational Background
}

\end{titlepage}

\author{
Ratindranath Akhoury$^{a}$ and Alexander D. Dolgov$^{b,c,d}$ 
}

\maketitle

\begin{center}
$^{a}$Department of Physics and Michigan Center for Theoretical Physics, \\ University of Michigan, Ann Arbor, Michigan 48109\\ 
$^{b}$Istituto Nazionale di Fisica Nucleare, Sezione di Ferrara,
       I-44100 Ferrara, Italy\\
$^{c}$Dipartimento di Fisica, Universit\`a degli Studi di Ferrara,
       I-44100 Ferrara, Italy\\
$^{d}$Institute of Theoretical and Experimental Physics,
       113259, Moscow, Russia\\
\end{center}

\vspace{0.5cm}

\begin{abstract}

We argue that superluminal propagation in a gravitational field discovered 
by Drummond and Hathrell in the lowest order of perturbation theory remains 
intact in higher orders. The criticism of this result based on an exact calculation
of the one loop correction to the photon polarization operator in the Penrose plane 
wave approximation is not tenable. The statement that quantum causality is automatically
imposed by classical causality is possibly invalid due to the infrared nature of the 
same triangle diagram which also contributes to the quantum trace anomaly.

\end{abstract}



Almost 30 years ago Drummond and Hathrell~\cite{drum} obtained a striking 
result that one-loop quantum corrections to Maxwell's equations in an external gravitational field lead to light propagation outside the light cone of the classical 
theory. The result depends upon the light ray direction and polarization state,
but still in some cases superluminal propagation is possible.
The effect owes its existence to the famous triangle diagram with an electronic loop and two photon 
and one graviton legs. On the other hand, a similar (box) diagram describing photon 
propagation in external electromagnetic field also leads to quantum distortion of
the light cone but, in this case, safely inside the classical one~\cite{adler}. 

This discovery of the possibility of super-luminal propagation is cause of serious concern if quantum effects could break causality. We recall here that the same triangle diagram
gives rise to the conformal anomaly and a similar one produces the chiral anomaly.
Logically there are several possibilities. Since the effect was found in the
lowest order of perturbation theory, higher order corrections might destroy it. The latter
can be either higher order corrections in electromagnetic interactions or higher orders
in gravity, or both. Another possibility is that the effect is real and motion faster than
light is possible but causality is not violated. There is no consensus in the literature
about the problem and all possible opinions are present, 
{\bf including causality violation}.

First, it is important to recall that formally the result of ref.~\cite{drum} is the {\it low} energy effective 
Lagrangian for photons in an external gravitational field (with an account of the lowest 
order quantum corrections). More specifically, in ref.~\cite{drum}, the 
the leading non-trivial term in an expansion of the quantum effective action in 
powers of ${q^2 \over m_e^2}$ ($q$ and $m_e$ are, respectively, the external momenta at the gravitational vertex and the electron mass) 
is kept. On the other hand, causality is determined by the velocity of the wave front, 
i.e. by the refractive index at infinitely
large photon frequency, see e.g.~\cite{mand}. The refractive index may exceed unity
at some finite frequency, $n(\omega) > 1$, and this would lead to group velocity
exceeding the speed of light. Such examples are known in media with anomalous dispersion.
However, if $n(\omega \rar \infty) \rar 1$, the speed of the wave front would be equal
to the speed of light and causality would be preserved. We will argue in what follows
that the result of ref.~\cite{drum} is true for any value of the photon frequency and
so is potentially dangerous for causality. It is worth noting that the faster than light
propagation found in vacuum with boundaries due to Casimir effect~\cite{casmr} is true
only for finite photon frequency, while at asymptotically large $\omega$, Casimir effect,
as is easy to see, disappears and thus super-luminal propagation due to the Casimir effect
is analogous to anomalous dispersion and does not violate causality. The same effect
is true in the Heisenberg-Euler~\cite{geisen} case:  the amplitude corresponding
to the box diagram describing 
light propagation in an external electric or magnetic field vanishes when 
$\omega \rar \infty$ and the speed of the wave front tends to the speed of light. Thus even 
if radiative corrections to light propagation in external electric or magnetic fields
lead to faster than light group velocity (we know that this is not the case
for propagation in external electromagnetic field~\cite{adler}),
the signal velocity would be bounded by the speed of light. The correction to the 
refractive index originating from the triangle diagram is a unique example which
does not vanish in the limit of $\omega \rar \infty$. It is simply because the three-leg
amplitude after extracting out spin factors can depend only on the momenta squared
of the participating particles~\cite{ibk}. In our case they are the momentum transferred
to the gravitational field, $q^2$ and $k_1^2$, $k_2^2$, where, $k_1, k_2$ are the momenta of the
propagating photons. Let us consider the imaginary part of the triangle diagram, thus the photon momenta can be taken to be on the mass-shell,
$k^2_1 = k_2^2 =0$. As discussed in~\cite{ad-vz} and in all generality in~\cite{coleman}, the kinematics of the triangle diagram (with spin factors extracted) is such that it has an imaginary part only at $q^2=0$, i.e., the imaginary part is proportional to $\delta(q^2)$, {\bf independent of the photon frequencies}. This is the origin of the statement that the anomaly is an infrared effect. In any case, the triangle diagram with spin factors factored out is independent of he photon frequency.
Next we note that the refractive index is known to be proportional to
the forward scattering amplitude and so in the lowest order in gravity 
$q^2=0$~\footnote{To calculate the refractive index as a function of coordinates,
we need to take the Fourier transform of the scattering amplitude
and obtain $n \sim 1/r^4$, which corresponds to small but non-zero momentum transfer.
The effect of that can be taken into account by consideration of the photon motion
along the geodesics.}. Since the triangle diagram has no dependance on the photon frequency,  the value of the refractive index found from the triangle 
graph is true for any $\omega$ and hence its low $\omega$ value is the same as
the value at $\omega \rar \infty$.  This leads 
to the speed of wave front propagation different from unity. 

As was argued in ref.~\cite{drum}, a once subtracted dispersion relation for 
refractive index
\be
n (\omega)  = n(0) +
\frac{2\omega }{\pi}\, \int d\omega'\, \frac{{\cal I}m\, n}{\omega - \omega'}, 
\label{disp-n}
\ee
allows us to conclude that $n(\infty) \leq n(0)$ if its imaginary part is non-negative,
${\cal I}m\, n \geq 0$. If this were true, the superluminal velocity found for small
$\omega$ would remain true for front propagation as well. However, it was explicitly 
shown in ref.~\cite{dol-khrip} that due to light focusing in a gravitational field, 
${\cal I}m\, n(\omega)$ may be negative:
\be
{\cal I}m\, n = \frac{\lambda}{2\omega^2} \, R_{0123},
\label{Im-n}
\ee 
where $\lambda$ is the photon helicity and $R_{0123}$ is the proper component of 
the Riemann tensor in the light cone coordinates, where the third coordinate
runs along the light ray geodesics. The conclusion that ${\cal I}m\, n$ may be
negative was later repeated by other authors, though some of them claimed that the
effect may be connected with particle production by the photon in gravitational field.
This effect is clearly negligible.

Thus dispersion relations cannot be used to prove that $n(\infty) \leq n(0)$ but 
nevertheless the problem persists because, as is mentioned above, the refractive index 
calculated from the triangle diagram does not depend on $\omega$. However, the loophole
remains that the higher order contributions of perturbation theory may cancel the lowest
order term. In what follows we analyze the structure of the perturbative expansion
and conclude that higher order corrections cannot remove the lowest order effect. Recently several papers~\cite{hol-shore}, have appeared where the authors state that in the limit
of the Penrose plane wave approximation one can calculate the electronic one-loop 
correction to photon propagation exactly and not just in the first order
in gravitational field. According to ref.~\cite{hol-shore} the exact result
does not allow for light to propagate outside the light cone. 
However, as we show below, the metric in the Penrose limit is too simple to allow for 
superluminal propagation even in the lowest order of perturbation theory, 
so going to this limit is essentially ``throwing the baby out with the bath water'' and the problem remains unsolved.

One may hope that despite superluminal propagation causality would not be violated,
as was argued in ref.~\cite{shore-causa}. However in ref.~\cite{dol-nov} an
explicit thought experiment was described which would surely lead to causality
violation if the law of transformation from one reference frame to another is 
known. Since the form of the superluminal radiative correction is clearly
Lorentz invariant (with an account of trivial complications due to space-time
curvature) see below eq. (\ref{L-eff}), the rules of the game are well defined and 
the transformation from one reference frame to another is determined. 
The arguments of ref.~\cite{dol-nov} are exactly the same as those used
to prove the well known statement that ``normal tachyons'' do break causality.
There are of course a few subtle points which make the problem somewhat
different but they can all be addressed with a precise treatment. First, one has
to take into account the effects of the nontrivial metric of General Relativity, namely 
time delay and distance variation. This can be 
easily done with a proper choice of the
coordinate system which can be taken along the trajectory as
\be 
ds^2 = A^2(l) dt^2 - B^2 (l) dl^2
\label{ds-traject}
\ee
In this coordinate frame it is easy to determine the analogue of the Lorentz
transformation in curved space-time. 

According to the results discussed here the photons have a position dependent 
velocity which may change from subluminal to superluminal and vice versa.
For the violation of causality it is sufficient that there are objects moving faster
than light somewhere in space and not necessarily everywhere. 

An apparent breakdown of Poincare invariance related to a fixed gravitating center,
e.g. a black hole, is not of primary importance because a ``time machine'' can be
made with two black holes moving in the opposite directions, each of them
creating superluminal motion in its vicinity. A source of photons is supposed to move
together with one of those two black holes and a detector is attached to another black 
hole. The detection of the photon near the second black hole switches on the source
of radiation which sends a photon back to the first black hole. The latter can come
back before the first photon was emitted by the source attached to the first black 
hole. The gravitational field of the second black hole, which quickly moves in the rest 
frame of the first black hole, would be noticeably different from the field in the
rest frame. In particular, it would be concentrated in a very narrow plane and the 
components of the Riemann tensor could be much larger than those in the original frame. 
However, as is argued in ref.~\cite{dol-nov}, this modification would not noticeably
effect the tachyon detection. All the technical details can be found in that paper.

Thus we are led to the troubling conclusion that in this problem, superluminal propagation 
quite likely leads to the quantum breaking of causality unless some, as yet unknown way, 
to eliminate the effect is found.

Another general argument forbidding superluminal propagation was presented
in reference~\cite{dub}. The authors argued that the causal structure of the
classical theory implies causality in the quantum field theory, i.e. the vanishing of 
the commutators of operators corresponding to local observables outside of the
light cone, if the corresponding classical Poisson brackets vanish. However, the authors themselves noted that the properties of classical theory might be distorted at the quantum level
due to anomalies, for example, the chiral or trace anomaly. Their an argument could be
possibly valid if anomalies are purely ultraviolet effects and thus  
modify the theory only by the addition of localized terms in position space, 
which cannot lead to the quantum distortion of the speed of light. However, as is
well known~\cite{ad-vz},~\cite{coleman}, and as mentioned above,  the chiral anomaly leads to an infrared modification of the theory creating an infrared pole $1/q^2$ in the t-channel, where $q$ is the
momentum transferred to the axial vertex. A similar infrared pole appears also
in the trace anomaly of the electromagnetic vertex in a gravitational 
field~\cite{q2-trace}. The triangle diagram which leads to the electromagnetic trace anomaly
is exactly the same as the one which leads to superluminal light propagation but
taken in a different kinematical region (for the anomaly calculation, the kinematic 
regime of interest is $q^2 > m_e^2$). This triangle diagram, as discussed earlier has an imaginary part only at $q^2=0$, independent of the photon frequency, which clearly brings out its infrared character. In this region the infrared nature of this diagram manifests itself 
as the singularity at $m_e =0$. Let us stress once more
that as is discussed in detail below, this is the only diagram which is singular at $m_e=0 $. All other
diagrams contain instead factors of $1/\omega$. Keeping this in mind, one should take the 
general statements about the equivalence of classical and quantum causality with 
great caution. We will return to this point at the end of this paper and present another argument to support this.

Moreover, as we explicitly demonstrate later,
an analysis of the perturbative expansion shows that higher order terms are
unable to remove the possibility of superluminal propagation. There might be of course some 
unknown non-perturbative effects forcing the refractive index to be less or equal to
unity. Though we do not see any particular reason for such terms to be
present, they, as ``Deus ex machina''(original Greek:``apo mechanes teos'') could easily solve all possible problems.   
 
We next turn to the criticism of the results of ref.~\cite{drum} 
as discussed in ~\cite{hol-shore}. 
 To set the stage, we first review some relevant arguments of ref.~\cite{drum}.
The radiative corrections to the Maxwell Lagrangian in a curved space-time background,
found in ref.~\cite{drum}, have the form:
\be
L_{eff} = -\frac{1}{4}\, F_{\alpha\beta} F^{\alpha\beta} 
+ \frac{\alpha}{m^2}\left( a_1 R F_{\alpha\beta} F^{\alpha\beta} + 
a_2 R^{\mu\nu} F_{\mu\alpha} F_\nu ^{\alpha} +
a_3 R^{\mu\nu\alpha\beta} F_{\mu\nu} F_{\alpha\beta} \right),
\label{L-eff}
\ee 
where the constant coefficients are $a_1= -1/144\pi$, 
$a_2=13/360\pi$, $a_3= -1/360\pi$, $\alpha = 1/137$
is the fine structure constant, and $m$ is the electron mass.

For light propagation in vacuum, outside gravitating bodies, where 
$R=0$ and $R^{\mu\nu} = 0$, only the last term proportional to the 
Riemann tensor makes non-vanishing contribution.
In particular, in the Schwarzschild geometry one polarization
state of the photon propagates, as expected, inside the light cone, while
the other polarization propagates outside it. The light cone is
defined by light propagation without quantum corrections, i.e. by the
classical Maxwell equation in curved background:
\be
D_\mu F^\mu_\nu = 0,
\label{DF}
\ee
where $D_\mu$ is the covariant derivative in the corresponding space-time.  

The coupling to the Riemann tensor, $R^{\mu\nu\alpha\beta}$, modifies 
the highest derivative terms in the equation of motion for $F_{\mu\nu}$ :
\be
D_\mu F^\mu_{\,\,\,\,\,\nu} - 
4a_3 D_\mu\left(R^{\mu}_{\,\,\,\,\,\nu\alpha\beta} F^{\alpha\beta}
\right) = 0.
\label{DRF}
\ee
To see this explicitly we take the isotropic metric:
\be
ds^2 = f dt^2 - g dr^2,
\label{ds-sch}
\ee
where 
\be
f = \left(\frac{1-r_g/4r}{1+r_g/4r}\right)^4,\,\,\,
g = \left( 1+ r_g/4r\right)^4
\label{f-g}
\ee
and $r_g = 2M /m_{Pl}^2$ is the gravitational radius of the source with mass $M$.

Assuming that light propagates along $z$ and only the transverse components of 
potential $A_j$ with $j=x,y$ are non-zero, we find that the second term in
eq.~(\ref{DRF}) contributes to higher derivative terms as
\be
-\partial_i\partial_j f\,(\partial_t^2 A_j) + \left(\partial_i\partial_j g+
\delta_{ij} \partial_z g\right) \partial_z^2 A_j.
\label{dt-dz}
\ee
It is easy to check that the distortion of the coefficient in front of the
time derivative is different
from that of the space derivative and as a result light does not propagate 
along the classical light cone. More detailed analysis shows that one polarization
state, as we mentioned above, goes outside cone, while the other one goes inside.
The effect changes sign after the light ray passes the minimal distance to the 
gravitating body (impact parameter). After that the polarization which propagated
faster than light slows down to subluminal velocity and vice versa for another
polarization.

Let us now  repeat the same exercise in the Penrose plane wave limit 
(PPW)~\cite{penrose} and
we will see that the second time and space derivatives are modified equally and
so in this approximation the light cone is not modified at all. A good description of
PPW can be found in the lectures~\cite{blau}. As is shown there, the Schwarzschild
metric in this limit in terms of normal Fermi coordinates~\cite{fermi} 
along the null geodesics takes the form:
\be
ds^2 = 2 du dv + F(u) (x^2-y^2) du^2 + dx^2 + dy^2,
\label{sch-PPW}
\ee
where $u$ and $v$ are the the light cone coordinates;
roughly speaking $u=t+z$ and $v = t-z$, and $x$ and $y$ are orthogonal to this
geodesics. The function $F(u)$ is essentially the Riemann tensor along this
geodesics, $F(u) \sim M/[m_{Pl}^2 r(u)^3]$, and $r(u)$ is the radial coordinate
along the geodesics.
The metric is so simple in this limit that the Riemann tensor has only one 
nonvanishing component, see e.g. eq. (2.62) of ref.~\cite{blau}:
\be 
R_{uiuj} = diag \left[-F(u), F(u) \right]
\label{R-uiuj}
\ee
Due to such simplicity of the space-time it may be possible to find one loop
corrections to the photon Green's function exactly for an external gravitational
field as is stated in ref.~\cite{hol-shore}. However, we think that this
is unnecessary because it is easy to see that superluminal propagation is
absent in the PPW limit. Moreover, we will see below that in the Penrose plane wave 
background, light propagates along the classical light cone at least in the lowest
order in gravity. According to the arguments presented later on, higher orders
in the gravitational field do not lead to superluminal propagation. Thus, if the the 
effect is absent in the lowest order it should be absent to all orders in a perturbative expansion.

The effective Lagrangian (\ref{L-eff}) has the same form in any 
coordinates, so we can easily write the Maxwell equations taking into account the
radiative corrections in the PPW limit. To this end we obtain for equation (\ref{DRF})
in the metric (\ref{sch-PPW}):
\be
\partial_u\partial_v A_k + \frac{g^{vv}}{2 + 4 a_3 R_{ukuk}}\,\partial_v^2 A_k =0.
\label{du-dv}
\ee
There is no summation over $k$ in this equation. 
Here $g^{vv} = -F(u)(x^2-y^2)$ and vanishes on the 
light geodesics. In the lowest order in gravitational field we can neglect
the term $4 a_3 R_{ukuk}$ which is induced by the radiative correction and
conclude that 
the equation describing quantum corrected light propagation 
in this metric essentially coincides with the classical 
Maxwell equation and light propagates along the classical geodesics. 

There is one more reason to expect that the PPW limit should be taken with care.
This limit corresponds to the reference frame moving with the speed 
asymptotically approaching the speed of light. In this limit the Schwarzschild 
gravitational field turns into delta-function shell orthogonal to the light
ray. The original Riemann tensor behaves as $R \sim 1/(x^2+y^2 +z^2)^{3/2}$.
At the transition to the PPW limit $x$ and $y$ do not change but $z$ transforms as
$z' = (z-Vx)\gamma$, where $V$ is the speed of the PPL reference frame, $V\rar 1$,
and $\gamma = 1/\sqrt{1-v^2}$. So we see that indeed the Riemann tensor in this 
reference frame becomes non-zero in a narrow shell of size 
$\delta \sim (x^2+y^2)/\gamma \rar 0$. Since, as mentioned above, the correction
to the speed of light changes sign along the light trajectory it may become zero
when averaged with $\delta (z-Vt)$. So the limit $\gamma =\infty$ must be
accurately regularized.

The consideration of the lowest order contribution to the photon
propagation in external gravitational field does not exclude in principle
that at asymptotically large frequencies the higher order corrections may
become larger than the first order term and thus save the world from acausal
propagation. An analysis of the higher order contributions to the
refractive index was carried out in ref.~\cite{dol-nov} and here we extend and generalize it.

We would like to note that the expansion parameter for
the diagrams with several graviton legs at high frequencies, $\omega \gg m$, 
is $R/\omega^2$ but not $R \omega^2/ m^4$, as is claimed in ref.~\cite{hol-shore}.
Here $R$ is a generic notation for any component of the Riemann  tensor.
The latter expansion parameter
is valid at low frequencies, and of course, at large frequencies the expansion
in terms of this parameter stops to be perturbative. Here we agree with
ref.~\cite{hol-shore}, however, a series in terms of $R/\omega^2$ is applicable 
at large $\omega$. The only exception to the expansion in terms of 
$R/\omega^2$ is the first ``triangle'' term which, as we noted above, is
frequency independent.

Dimensional counting shows that higher order corrections in the 
electromagnetic interactions are not dangerous. 
This result is practically evident because quantum electrodynamics is
known to be renormalizable in an external gravitational 
background~\cite{renorm-QED}. For example, the box diagram with two photon and
two graviton legs is easy to estimate and one can explicitly check
that the scattering amplitude generated by such a graph at high photon
frequencies vanishes as $1/s$, where $s = k_1 q_j$, $k_1$ is the momentum
of the initial photon (note that $k_1= k_2$) and $q_j$ is the momentum transferred 
to either graviton vertex. We see, as expected, that the high energy amplitude is not
singular at $m =0$.

Another class of diagrams includes additional external graviton legs.
The amplitude corresponding to such diagrams should contain terms of
the type $F^2 R^n/\mu^{2n} $, where $\mu$ is a factor of dimension of mass,
$F$ is the Maxwell tensor and $R$ is one of the curvature tensors. 
The contraction over the indices, not explicit here, is assumed. 
Such terms are evidently not dangerous. 

The problematic terms which increase with
the photon energy terms should contain derivatives, 
$ F (D^{2k} F) R^n /\mu^{2n+2k}$. The factor $\mu^{2n+2k}$ may contain the
electron mass, $m$ and some kinematical variables, as. e.g. $k_1 q_j$,
where $k_1$ is the photon momentum and $q_j$ is the momentum
transferred to the gravitational field in $j$-th vertex. In contrast to
the triangle graph considered in the previous section, where $\mu=m$ and
the result was singular for vanishing electron mass, more complicated
diagrams depend upon the real kinematical variables and are not singular
at $m=0$. Thus we expect that higher order corrections due to
additional couplings to the external gravitational field do not increase as
$\omega \rar \infty$. This result is a manifestation of the already
mentioned renormalizability of QED in an external gravitational field.

There exists another class of diagrams with two photon legs, several
gravitational legs, and some internal graviton exchange. Each internal
graviton line comes together with the factor of the inverse Planck
mass squared, $1/m_{Pl}^2$. Such diagrams are strongly ultraviolet
divergent due to the non-renormalizability of quantum gravity. Naively
they should be UV-cut-off at the Planck scale, $\Lambda_{UV} = m_{Pl}$.
So there would be the terms of the type $(\Lambda_{UV}/m_{Pl})^{2k}$,
where $k$ is an integer equal to the number of the internal graviton
lines. However, there could be also the terms proportional to 
$(\omega/m_{Pl})^{2k}$. Such terms may become dangerous at huge photon 
frequencies, $\omega \sim m_{Pl}$. To eliminate faster-than-light
propagation such higher order terms must be of the same magnitude as
the lowest order one, discussed above. If quantum gravity is really
non-renormalizable, then such terms are difficult to exclude. On the
other hand, if the ``great expectations'' of a finite or at least
renormalizable theory of gravity is to be realized, then it is natural
to expect that the scattering amplitudes do not increase with energy
and hence the higher order exchange of gravitons would not lead to any significant effects
even for very energetic photons. Thus the problem of super-luminal
propagation probably cannot be solved within the framework of a possible
future theory of gravity which has a well defined UV completion.

Let us return now to the earlier discussion about the statement that causal propagation
in the classical theory implies causality in quantum theory due to vanishing of
field commutators outside the light cone. We first discuss wave propagation 
in the classical theory and then turn to the problem of causality in quantum field 
theory. The classical wave equation for light (or any other massless wave)
propagation has the form:
\be
D^2 \phi \equiv (\partial_t^2 - \Delta ) \phi = J\,,
\label{dt-phi}
\ee
where $\phi$ describes the wave amplitude and $J$ is a source. The solution of this 
equation which respects the causality condition is given by
\be
\phi_0 = \int d^4 y G_R (x-y) J(y)\,,
\label{phi-0}
\ee
where $G_R$ is the retarded Green's function equal to:
\be
G_R(x) = \frac{1}{(2\pi)^4} \int_{C_R} \frac{dE d^3 p \exp (-ip x)}{E^2- p^2},
\label{G-R}
\ee
and the contour $C_R$ goes above the poles at $E = \pm p $. 

Now, if for some reason the equation of motion is modified to
\be
\tilde D^2 
\equiv (\partial_t^2- \Delta + \Gamma_{\mu\nu} \partial_\mu\partial_\nu ) \phi = J\,,
\label{tilde-dt-phi}
\ee
it is evident that the solution of this equation expressed through the
retarded Green's function $\tilde G_R$ of operator $\tilde D^2$ in the same
way as eq.(\ref{phi-0}) may propagate outside the cone
of eq. (\ref{dt-phi}) for some choice of function $\Gamma_{\mu\nu}$:
\be
\tilde \phi = \int d^4 y \tilde G_R (x-y) J(y)\,,
\label{phi-tilde}
\ee

On the other hand if we solve eq. (\ref{tilde-dt-phi}) perturbatively with respect to
$\Gamma_{\mu\nu}$ using the retarded Green's function of the unperturbed operator $D^2$,
we will find order by order that the signal propagates inside the old cone of 
eq.~(\ref{dt-phi}):
\be
\tilde \phi = \phi_0 - \int d^4 y G_R (x-y) 
\Gamma_{\mu\nu} \partial_\mu\partial_\nu \phi_0 (y) + ... \,,
\label{-tilde-phi-G-R}
\ee
where $\phi_0$ is given by eq. (\ref{phi-0}). 

Keeping this in mind let us turn now to quantum theory. The problem of causality
in quantum field theory is less trivial than that in classical 
physics, see e.g. ref.~\cite{causa-quant,peskin}. The
difficulty lies in imposing the positive energy (frequency) condition on the
Fourier decomposition of the particle wave function. Evidently it is impossible to 
confine a function in finite spatial region using positive frequencies only.
However, causality is not violated, almost by magic, due to the existence of 
particles which ``travel backward in time'' (antiparticles). 
{\bf Formally it looks as negative frequencies.} 
This effect essentially {\bf means} that the acausal Feynman Green's function is
substituted for the retarded causal Green's function in the description of 
signal propagation. More details can be found in ref.~\cite{causa-quant,peskin}. 

The usual proof of causality is based on the statement of the vanishing of 
the commutator of bosonic field operators (anticommutator for fermions) 
outside the light cone.
To make things clearer let us reproduce the arguments presented in
~\cite{peskin}. The commutator of the field $\phi$ at two different points is
equal to:
\be
\left[ \phi(x),\phi(y) \right] &=& 
\int \frac{d^3 p d^3 q}{(2\pi)^6 \sqrt{4E_pE_q}}
\left[\left(a_{\bf p} e^{-i px} + a_{\bf p}^\dagger e^{ip x}\right),
\left(a_{\bf q} e^{-i qx} + a_{\bf q}^\dagger e^{iq x}\right)\right] \nonumber\\
&=& \int\frac{d^3p}{2E_p (2\pi)^3} \left( e^{-ip(x-y)} - e^{ip (x-y)} \right)
= D(x-y) - D (y-x)\,,
\label{commutator}
\ee
where $E_p = \sqrt{p^2 + m^2}$, $a^\dagger_p$ and $a_p$ are the creation and
annihilation operators of quanta with momentum $p$ and 
\be
D(x-y) = \langle 0 | \phi(x) \phi(y) |0\rangle 
\label{D-of-x-y}
\ee 
is the Lorentz invariant amplitude of particle propagation from the
space-time point $x$ to the point $y$.
It can be seen that for a time-like interval $(x-y)$ and large time differences
this function oscillates. In the reference frame where the 4-vector $(x-y)$ 
has only a time component $D(x-y) \sim \exp (-i mt) $. 
For a space-like interval, $(x-y) = (0, {\bf r})$, function $D(x-y)$ exponentially
decreases remaining non-zero outside the light cone,
$D(0,r) \sim \exp (-mr)$. However, this does not imply causality violation
which is determined by the commutator (\ref{commutator}).  
It is easy to check that this commutator indeed does vanish for a space-like interval.
According to ~\cite{causa-quant}, the all important second term in the 
expression (\ref{commutator}) comes from particles propagating backward in
time. We have to stress here that the proof is valid for free field operators. 
The statement of the vanishing of the commutator remains true
for the interacting fields within the framework of perturbation theory. However, as we 
have seen above, even in classical field theory with the distorted equation of motion,
which surely breaks causality, the perturbative solution remains causal. So we should
expect the same phenomenon in quantum physics. Perturbative proof of causality is
violated by the solution of the exact equation of motion if the latter breaks causality.

In the case under consideration here the classical equation of motion for 
photons (\ref{DRF}) is surely acausal. If we solve this equation perturbatively
with respect to the second term proportional to the Riemann tensor, we will
find a causal solution. However, as we have demonstrated above, a perturbative solution 
is not sensitive to causality violation. 

At this stage it is worth noting that one should distinguish between two kinds 
of perturbation theory which we use here. The first is the perturbative calculations
of the one loop correction to light propagation in gravitational field, which modifies
the causal Maxwell equations of motion. As we argued above, this perturbative correction 
is the only one which survives at asymptotically high frequencies. In this sense this
perturbative result is exact. As for the solution of the perturbed equation (\ref{DRF}),
it should be done exactly, non-perturbatively. 
In other words, the classical theory with the equation of motion (\ref{DRF}) 
is acausal and so should be the quantum theory obtained by a quantization of this
classical one. 

Thus, we have failed to see if and how superluminal propagation can be avoided.
An explicit analysis of the perturbative expansion shows that the higher
order terms do not change the lowest order result and the existing general proofs in the literature,
to the contrary, appear to suffer from serious drawbacks.
 We conclude that in the context of a local quantum field theory, causality violation induced by 
quantum effects appears to be realistic. In a future publication we hope to discuss the possible effects of superluminal photons, for example, their capture by black holes.

We thank Sergei Dubovsky and Alexander Vikman for discussions and for their very useful critical comments on the manuscript. A.D. thanks the Michigan Center for Theoretical Physics for 
hospitality during the time when this work was done.


\end{document}